\documentclass[11pt]{spie}  
\usepackage[]{color,graphicx}
\usepackage{setspace}
\usepackage{tocloft}

\title{The Transiting Exoplanet Survey Satellite} 

\title{The Transiting Exoplanet Survey Satellite}

\author{\small
George~R.~Ricker\supscr{a},
Joshua~N.~Winn\supscr{a},
Roland~Vanderspek\supscr{a},
David~W.~Latham\supscr{b},
G\'{a}sp\'{a}r~\'{A}.~Bakos\supscr{c},
Jacob~L.~Bean\supscr{d},
Zachory~K.~Berta-Thompson\supscr{a},
Timothy~M.~Brown\supscr{e},
Lars~Buchhave\supscr{b,f},
Nathaniel~R.~Butler\supscr{g},
R.~Paul Butler\supscr{h},
William~J.~Chaplin\supscr{i,j},
David~Charbonneau\supscr{b},
J{\o}rgen~Christensen-Dalsgaard\supscr{j},
Mark~Clampin\supscr{k},
Drake~Deming\supscr{l},
John~Doty\supscr{m},
Nathan~De~Lee\supscr{n,o},
Courtney~Dressing\supscr{b},
E.~W.~Dunham\supscr{p},
Michael~Endl\supscr{q},
Francois~Fressin\supscr{b},
Jian~Ge\supscr{r},
Thomas~Henning\supscr{s},
Matthew~J.~Holman\supscr{b},
\mbox{Andrew~W.~Howard\supscr{t},}
Shigeru~Ida\supscr{u},
Jon~M.~Jenkins\supscr{v},
Garrett~Jernigan\supscr{w},
John~Asher~Johnson\supscr{b},
Lisa~Kaltenegger\supscr{s}, 
Nobuyuki~Kawai\supscr{u},
Hans~Kjeldsen\supscr{j},
Gregory~Laughlin\supscr{x},
Alan~M.~Levine\supscr{a},
Douglas~Lin\supscr{x},
Jack~J.~Lissauer\supscr{v},
Phillip~MacQueen\supscr{q},
Geoffrey~Marcy\supscr{y}, 
P.~R.~McCullough\supscr{z,aa},
Timothy~D.~Morton\supscr{c},
Norio~Narita\supscr{bb},
Martin~Paegert\supscr{o},
\mbox{Enric~Palle\supscr{cc},}
Francesco~Pepe\supscr{dd},
Joshua~Pepper\supscr{o,ee},
Andreas~Quirrenbach\supscr{ff},
S.~A.~Rinehart\supscr{k},
Dimitar~Sasselov\supscr{b},
Bun'ei~Sato\supscr{u},
Sara~Seager\supscr{a},
Alessandro~Sozzetti\supscr{gg},
Keivan~G.~Stassun\supscr{o,hh},
Peter~Sullivan\supscr{a},
Andrew~Szentgyorgyi\supscr{b},
Guillermo~Torres\supscr{b},
Stephane~Udry\supscr{dd},
Joel~Villasenor\supscr{a}
\skiplinehalf
\footnotesize
\supscrsm{a}Massachusetts Institute of Technology;  
\supscrsm{b}Harvard-Smithsonian Center for Astrophysics;  
\supscrsm{c}Princeton University;
\supscrsm{d}University of Chicago;  
\supscrsm{e}Las Cumbres Observatory Global Telescope;  
\supscrsm{f}University of Copenhagen, Denmark;  
\supscrsm{g}Arizona State University;  
\supscrsm{h}Department of Terrestrial Magnetism, Carnegie Institute of Washington;  
\supscrsm{i}University of Birmingham, UK;  
\supscrsm{j}Stellar Astrophysics Centre, Aarhus University, Denmark;  
\supscrsm{k}NASA Goddard Space Flight Center;  
\supscrsm{l}University of Maryland;  
\mbox{\supscrsm{m}Noqsi Aerospace, Ltd.;}  
\supscrsm{n}Northern Kentucky University;  
\supscrsm{o}Vanderbilt University;  
\supscrsm{p}Lowell Observatory;
\supscrsm{q}McDonald Observatory;  
\supscrsm{r}University of Florida;  
\supscrsm{s}Max-Planck-Institut f\"{u}r Astronomie, Heidelberg, Germany;  
\supscrsm{t}University of Hawaii;
\supscrsm{u}Tokyo Institute of Technology, Japan;  
\mbox{\supscrsm{v}NASA Ames Research Center;}
\supscrsm{w}Space Science Laboratory, University of California, Berkeley;  
\mbox{\supscrsm{x}UCO/Lick Observatory;}
\supscrsm{y}University of California, Berkeley;
\supscrsm{z}Space Telescope Science Institute;  
\mbox{\supscrsm{aa}Johns Hopkins University;}
\supscrsm{bb}National Astronomical Observatory of Japan;
\supscrsm{cc}Instituto de Astrofisica de Canarias, Spain;  
\supscrsm{dd}Observatoire de Gen\`{e}ve, Switzerland;
\supscrsm{ee}Lehigh University;
\mbox{\supscrsm{ff}Landessternwarte, Zentrum f\"{u}r Astronomie der Universit\"{a}t
Heidelberg, Germany;}
\mbox{\supscrsm{gg}INAF--Osservatorio Astrofisico di Torino, Italy;}
\supscrsm{hh}Fisk University
\normalsize
}

\cftpagenumbersoff{figure}
\cftpagenumbersoff{table} 

\def\ltsima{$\; \buildrel < \over \sim \;$}
\def\lsim{\lower.5ex\hbox{\ltsima}}
\def\gtsima{$\; \buildrel > \over \sim \;$}
\def\gsim{\lower.5ex\hbox{\gtsima}}
\def\tess{{\it TESS}\ }

\def\skiplinehalf{\medskip\\}

\def\supscr#1{\raisebox{0.8ex}{\small #1}\hspace{0.05em}}  
 
\def\supscrsm#1{\raisebox{0.8ex}{\footnotesize #1}\hspace{0.05em}}  

\begin{document} 
\maketitle 

\begin{abstract}
  The Transiting Exoplanet Survey Satellite ({\it TESS}) will search
  for planets transiting bright and nearby stars. \tess has been
  selected by NASA for launch in 2017 as an Astrophysics Explorer
  mission. The spacecraft will be placed into a highly elliptical
  13.7-day orbit around the Earth. During its two-year mission, \tess
  will employ four wide-field optical CCD cameras to monitor at least
  200,000 main-sequence dwarf stars with $I_C\approx 4$-13 for
  temporary drops in brightness caused by planetary transits. Each
  star will be observed for an interval ranging from one month to one
  year, depending mainly on the star's ecliptic latitude. The longest
  observing intervals will be for stars near the ecliptic poles, which
  are the optimal locations for follow-up observations with the {\it
    James Webb Space Telescope}. Brightness measurements of
  preselected target stars will be recorded every 2~min, and full
  frame images will be recorded every 30~min. \tess stars will be
  10-100 times brighter than those surveyed by the pioneering {\it
    Kepler} mission. This will make \tess planets easier to
  characterize with follow-up observations. \tess is expected to find
  more than a thousand planets smaller than Neptune, including dozens
  that are comparable in size to the Earth. Public data releases will
  occur every four months, inviting immediate community-wide efforts
  to study the new planets. The \tess legacy will be a catalog of the
  nearest and brightest stars hosting transiting planets, which will
  endure as highly favorable targets for detailed investigations.
\end{abstract}

\keywords{exoplanet, extrasolar planet, photometry, satellite, transits}

{\noindent \footnotesize{\bf Address all correspondence to}: Dr.\
  George Ricker, Massachusetts Institute of Technology, 77
  Massachusetts Avenue, Room 37-501, Cambridge, MA 02139-4307; Tel: +1
  617-253-7532; E-mail:~{\tt grr@space.mit.edu}


\begin{spacing}{1}   


\section{INTRODUCTION}
\label{sec:intro}

The study of exoplanets---planets outside our Solar System---is one of
the most exciting and rapidly advancing fields of science. Especially
valuable are systems in which a planet's orbit carries it directly
across the face of its host star. For such a ``transiting'' planet, it
is possible to determine the planet's mass and radius, its orbital
parameters, and its atmospheric properties\cite{winn11}.

Of particular interest are planets with sizes between those of the
Earth and Neptune. Little is known about them, because there are no
examples in the Solar System. NASA's {\it Kepler} mission
revolutionized exoplanetary science by revealing that such planets are
abundant\cite{borucki11,fressin13}, and seem to have a wide range of
compositions\cite{marcy14} and interesting orbital
configurations\cite{lissauer11}. However, most of the {\it Kepler}
stars are too faint for detailed follow-up observations.

The Transiting Exoplanet Survey Satellite, or {\it TESS}, will take
the next logical step after {\it Kepler} by searching the nearest and
brightest stars for transiting planets. The primary objective of \tess
is to discover hundreds of transiting planets smaller than Neptune,
with host stars bright enough for follow-up spectroscopy to measure
planetary masses and atmospheric compositions. This objective will be
achieved by conducting a two-year all-sky survey, during which
differential time-series photometry will be performed for hundreds of
thousands of stars. The main value of \tess data will not be
statistical completeness, but rather the relative ease of following up
on discoveries with current and planned instruments.

This paper summarizes the {\it TESS} mission: its history
(\S~\ref{sec:history}), design considerations (\S~3), payload (\S~4),
spacecraft (\S~5), orbit (\S~6), operations (\S~7), anticipated
results (\S~8), institutional partners (\S~9), and broader context
(\S~10). Since the \tess mission is still in Phase B as of writing,
the information presented here is provisional and reflects the design
as of mid-2014.

\section{HISTORY} 
\label{sec:history}

The initial discussions leading to the \tess concept began at the
Massachusetts Institute of Technology (MIT) and the Smithsonian
Astrophysical Observatory (SAO) in late 2005. In early 2006, a
proposal for a Mission of Opportunity was submitted to NASA to use the
optical camera of the High Energy Transient Explorer\cite{ricker03} to
perform a survey for transiting planets. This proposal was not
selected.

Later in 2006 and early 2007, \tess was reformulated as a standalone
small mission. Seed funding was obtained from private individuals, the
Kavli Foundation, Google, MIT, the Smithsonian Astrophysical
Observatory, and the NASA Ames Research Center, but attempts to raise
the funds required for a full mission were not successful.

In 2008, \tess was reconceived as a Small Explorer Mission (SMEX), a
NASA program with a cost cap of \$100m. The \tess SMEX proposal was
funded for a Phase A study during 2008-2009, but was not chosen to
proceed into Phase B.

During the next two years, the \tess concept was gradually refined,
and a new proposal was submitted to the NASA Explorer program in 2011.
The larger cost cap (\$200m) enabled several key improvements,
particularly the use of a high-Earth elliptical orbit, which provides
a more stable platform for precise photometry than the low-Earth orbit
that had been the basis of all the previous proposals. \tess was
selected by NASA for a Phase A study as an Explorer mission in Fall
2011, and proceeded to Phase B in April 2013.

\section{DESIGN CONSIDERATIONS} 
\label{sec:design}

This section describes some of the considerations that led to the
current design of the \tess mission.

\subsection{Sky coverage}

Although many transiting planets have been found, relatively few of
them orbit stars bright enough to enable follow-up measurements of
planet masses and atmospheres. Therefore, a prime objective of {\it
  TESS} is to monitor bright stars. Since the brightest stars are
nearly evenly distributed over the entire sky, this desire led in the
direction of an all-sky survey.

\subsection{Period sensitivity}
 
Ideally the mission would detect planets with a broad range of orbital
periods, with $P_{\rm min}$ near the Roche limit of a few hours, and
$P_{\rm max}$ of a year or longer. However, the choice of $P_{\rm
  max}$ determines the mission duration and thereby has a
disproportionate influence on cost. Furthermore, transit surveys are
strongly and inherently biased toward short periods.\footnote{As shown
  in refs.\ 7 and 8, in an idealized imaging survey limited only by
  photon-counting noise, the effective number of stars that can be
  searched for transiting planets of period $P$ varies as $P^{-5/3}$.}
Indeed, the period distribution of {\it Kepler}
detections\cite{borucki11} reaches a maximum at $\approx$10~days,
taking into account the period dependencies of both the transit
detection efficiency and the occurrence of planets. Hence, even a
value of $P_{\rm max}$ as short as 10~days should yield many
discoveries.

Choosing $P_{\rm max}$ as short as 10~days rules out the detection of
habitable-zone planets around Sun-like stars. However, it is possible
to gain access to habitable-zone planets around M stars if at least a
portion of the sky is observed with $P_{\rm max}\gsim
40$~days.\footnote{For an M0 star with $T_{\rm eff} = 3800$~K, the
  range of orbital distances receiving a bolometric insolation within
  a factor of two of the Earth's insolation corresponds roughly to
  $P=25$--75~days.}  And if there is a portion of the sky to be
searched more extensively, it would be advantageous for that portion
to coincide with the zones of longest continuous visibility with the
{\it James Webb Space Telescope} ({\it JWST}\,), which will likely be
the best telescope for follow-up studies of planetary
atmospheres. Those zones are centered on the ecliptic
poles.\cite{gardner06}

Together these considerations led to the choices of $P_{\rm
  max}\approx 10$~days in general, and $P_{\rm max}\gsim 40$~days for
as large an area as possible surrounding the ecliptic poles.

\subsection{Type of stars monitored}

While it would be interesting to survey all types of stars, it is
logical to concentrate on main-sequence dwarfs with spectral types F5
to M5. Evolved stars and early-type dwarfs are large, inhibiting the
detection of small planets. Dwarfs with spectral types earlier than F5
also rotate rapidly, broadening their spectral lines and preventing
precise radial-velocity monitoring.

On the other side of the spectral sequence, M dwarfs are especially
attractive targets. They are abundant: about three-quarters of the
stars in the solar neighborhood are M0-M5 dwarfs. They are relatively
unexplored for transiting planets, because they constituted only a
small minority of the {\it Kepler} target list. Furthermore, the
transit signal of a small planet is easier to detect for an M dwarf
than it would be for a larger star of the same apparent magnitude,
facilitating both planet discovery and follow-up observations with
{\it JWST} and other telescopes.

However, stars with spectral types later than M5 are rarer and
optically faint. They could be observed advantageously at
near-infrared wavelengths, but this would greatly increase the
mission's cost, complexity, and risk. Furthermore, planets transiting
the very latest-type stars can be detected with ground-based
instruments, as demonstrated by the MEarth survey.\cite{charbonneau09}
For these reasons, the F5-M5 range of spectral types was considered to
be most important for {\it TESS}.

\subsection{Detector and bandpass}

The best astronomical detectors in the optical band are silicon
charge-coupled devices (CCDs), which have excellent linearity and
dynamic range for bright objects, and a long spaceflight heritage. A
wide bandpass is preferred to reduce photon-counting noise. However,
very wide bands present challenges in constructing a wide-field and
well-focused optical system. Enhanced sensitivity to red wavelengths
is desired because small planets are more easily detected around small
stars, which are cool and red.

These considerations led to the choice of a 600-1000~nm bandpass. The
red end represents the red limit of silicon CCD sensitivity, and
the width of 400~nm was the largest practical choice for the optical
design. This bandpass is centered on the traditional $I_C$ band but is
much wider (see Fig.~1). It is comparable to the union
of the $R_C$, $I_C$, and $z$ bands.

\begin{figure}
\begin{center}
\begin{tabular}{c}
\includegraphics[height=7cm]{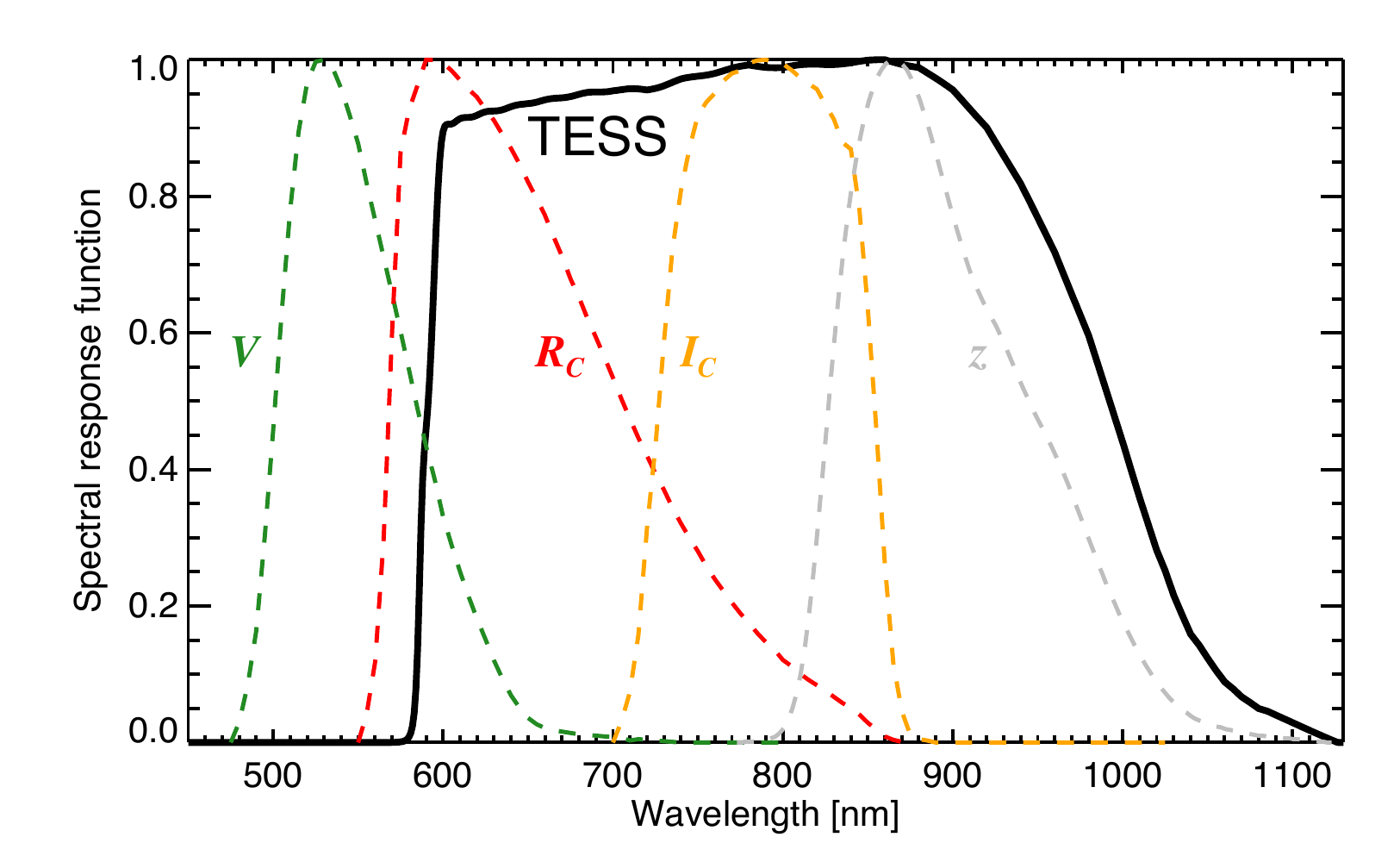}
\end{tabular}
\end{center}
\caption{ \label{fig:Bandpass} The TESS spectral response
  function (black line), defined as the product of the long-pass
  filter transmission curve and the detector quantum efficiency curve.
  Also plotted, for comparison, are the Johnson-Cousins $V$, $R_C$,
  and $I_C$ filter curves and the SDSS $z$ filter curve. Each of the
  functions has been scaled to have a maximum value of unity.}
\end{figure} 

\subsection{Number of stars monitored}

{\it Kepler}'s transit detection rate\cite{borucki11} for
``super-Earths'' (1.25--2~$R_\oplus$) with $P<10$~days was about
0.2\%.  This represents the product of the occurrence rate ($\sim$5\%)
and the geometric transit probability ($\sim$5\%). Therefore, \tess
needs to monitor at least 500 stars for each super-Earth
detection. The desire to find hundreds of such planets leads to the
requirement to monitor $\gsim$$10^5$ stars.

According to Galactic star count models\cite{girardi05}, the brightest
$10^5$ dwarf stars have a limiting apparent magnitude of $I_C\approx
10$. In fact the \tess design employs a larger catalog
($\gsim$$2\times 10^5$ stars) selected according to transit
detectability rather than strictly on apparent magnitude. This leads
to a limiting magnitude of $I_C \lsim 10$-13, depending on spectral
type. The faint M dwarfs, for example, have such small sizes that they
can still be usefully searched for small planets.

\subsection{Aperture size}

Stars must be monitored with sufficient photometric precision to
detect planetary transits. This leads to requirements on the
collecting area and other properties of the cameras, although the
derivation of those requirements is not straightforward because of the
wide variation in the characteristics of the transit signals.  As
reviewed in Ref.\ 1, the transit depth varies with the radii of the
star ($R_\star$) and planet ($R_p$),
\begin{equation}
\delta = (337~{\rm ppm})
\left( \frac{R_p}{2~R_\oplus} \right)^2
\left( \frac{R_\star}{R_\odot} \right)^{-2},
\end{equation}
and the transit duration ($T$) depends on the stellar mean density
($\rho_\star$) as well as the orbital period ($P$) and dimensionless
transit impact parameter ($b$),
\begin{equation}
T = (3.91~{\rm hr})
\left( \frac{\rho_\star}{\rho_\odot} \right)^{-1/3}
\left( \frac{P}{10~{\rm days}} \right)^{1/3}
\sqrt{1-b^2}.
\end{equation}
In these equations the fiducial case is a transiting planet with
$R_p=2~R_\oplus$ in a 10-day circular orbit around a Sun-like star
($R_\star = R_\odot$, $\rho=\rho_\star$).

Monte Carlo simulations were performed\cite{sullivan14} to establish
the requirements for detecting a certain number of transiting planets,
taking into account the likely distribution of stellar apparent
magnitudes and radii; the occurrence of planets of various sizes and
orbits; as well as noise due to sky background, instrumental readout,
and other sources. The results of this study, summarized in
\S~\ref{sec:results}, indicate that detecting hundreds of super-Earths
requires $\approx$50~cm$^2$ of effective collecting area (i.e., the
actual collecting area multiplied by all throughput factors including
transmission losses and quantum efficiency).  This corresponds to an
effective aperture diameter of $D\approx 10$~cm.

\subsection{Time sampling}

Since the typical transit durations are measured in hours, the time
sampling of the photometric observations should be substantially less
than an hour. The timescale of the partial transit phases (ingress and
egress) is $\sim$$T(R_p/R_\star)$, which is typically several minutes
for planets smaller than Neptune. Therefore, to avoid excessive
smearing of the partial transit signals, a time sampling of a few
minutes or shorter is preferred. For {\it TESS} the brightness
measurements of the preselected stars will be recorded every 2~min or
shorter, which is consistent with these requirements and also enables
asteroseismology (see \S~\ref{sec:asteroseismology}).

\subsection{Orbit}

The ideal orbit for the \tess spacecraft would allow for continuous
observations lasting for weeks, and would be stable to perturbations
over the multi-year duration of the survey. It would offer a
low-radiation environment, to avoid high trapped-particle fluxes and
resulting degradation of the CCDs and flight electronics. It would
also offer a stable thermal environment and minimal attitude
disturbance torques, to provide a stable platform for precise
photometry. The orbit would be achievable with a moderate $\Delta V$,
avoiding the need for a costly secondary propulsion unit. Furthermore,
to facilitate data transfer, the spacecraft would be close to the
Earth during at least a portion of the orbit. As explained in
\S~\ref{sec:orbit}, these requirements were met by choosing an
elliptical orbit around the Earth, with approximate perigee and apogee
distances of $17$ and $59~R_\oplus$, and a period of 13.7~days.

\section{PAYLOAD}
\label{sec:payload}

The \tess payload consists of four identical cameras and a Data
Handling Unit (DHU). Each camera consists of a lens assembly with
seven optical elements, and a detector assembly with four CCDs and
their associated electronics. All four cameras are mounted onto a
single plate. These components will now be discussed in turn.

\subsection{Lens Assembly}

\begin{figure}
\begin{center}
\begin{tabular}{c}
\includegraphics[height=8cm]{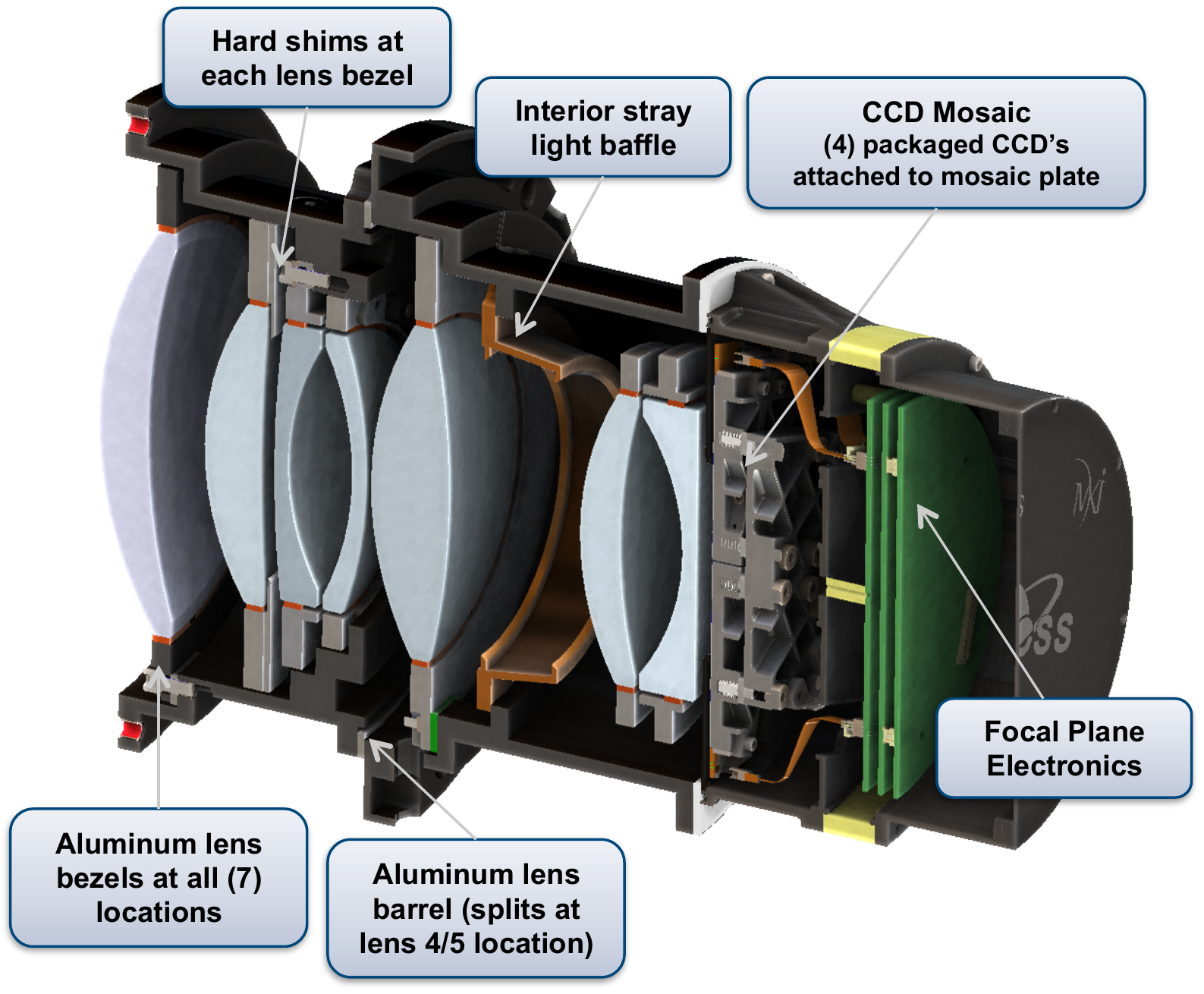}
\end{tabular}
\end{center}
\caption
{ \label{fig:Lens} 
Diagram of the lens assembly, CCD focal plane, and detector
electronics, for one of the four \tess cameras.}
\end{figure} 

\begin{figure}[b!]
\begin{center}
\begin{tabular}{c}
\includegraphics[height=6.5cm]{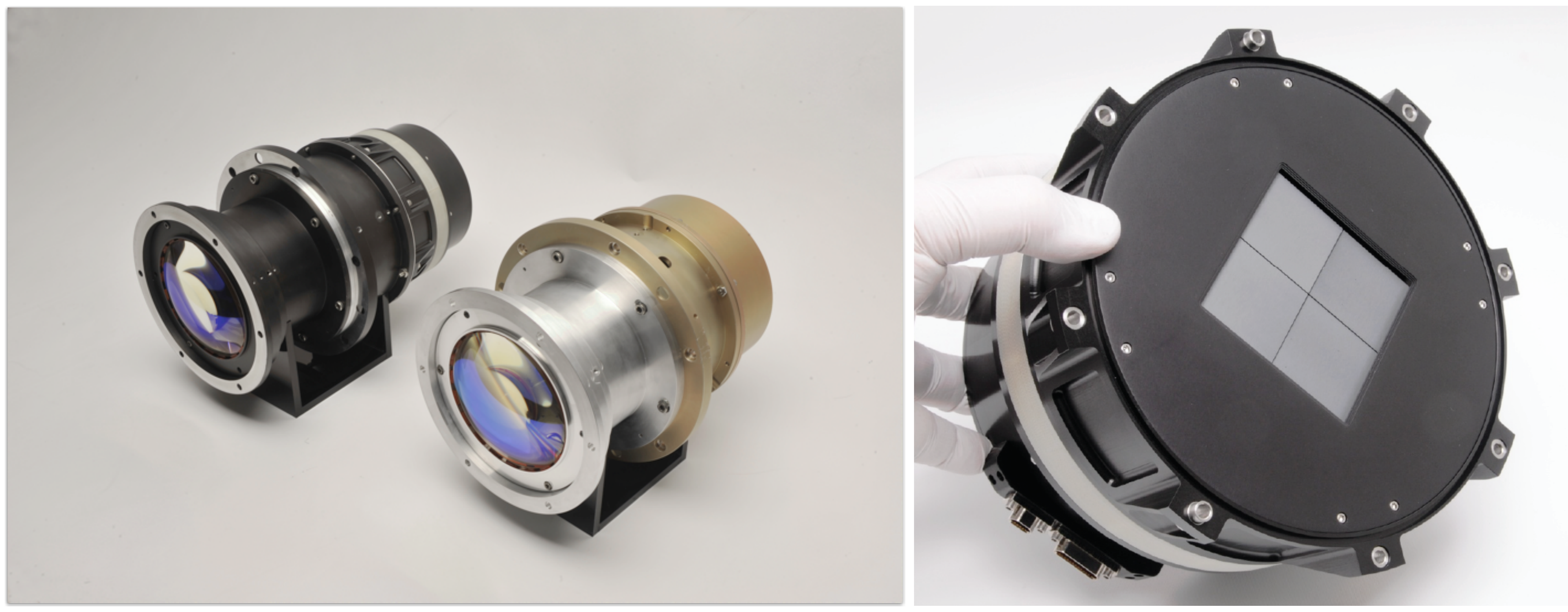}
\end{tabular}
\end{center}
\caption { \label{fig:PrototypeLens}
{\it Left.}---Two lens prototypes were constructed
  during Phase A. One was subjected to thermal
  vacuum testing at the operational temperature; the other
  was subjected to vibration testing.
{\it Right.}---The detector assembly of one of the prototype
lenses. The
frame-store regions of the CCDs are covered. }
\end{figure} 

Each of the four identical \tess lenses is an $f/1.4$ custom design
consisting of seven optical elements, with an entrance pupil diameter
of 10.5~cm (see Figs.~2-3 and Table~\ref{tbl:Lens}). For ease of
manufacture, all lens surfaces are spherical except for two mild
aspheres. There are two separate aluminum lens barrels that are
fastened and pinned together. All optical elements have antireflection
coatings. The surface of one element also has a long-pass filter
coating to enforce the cutoff at 600~nm. The red limit at 1000~nm is
set by the quantum-efficiency curve of the CCDs (see
Fig.~1).

Each lens forms a $24^\circ \times 24^\circ$ unvignetted image on the
four-CCD mosaic in its focal plane. The optical design was optimized
to provide small image spots of a consistent size across the field of
view (FOV), and produce undersampled images similar to those of {\it
  Kepler}\cite{gilliland11}.  At nominal focus and flight temperature
($-75^\circ$~C), the 50\% ensquared-energy half-width is 15~$\mu$m
(one pixel or 0.35~arcmin) averaged over the FOV. Each lens is
equipped with a lens hood, which reduces the effects of scattered
light from the Earth and Moon.
 
\begin{table}[t!]
  \caption{Characteristics of the \tess lenses. Ensquared
    energy is the fraction of the total energy of the point-spread
    function that is within a square of the given
    dimensions centered on the peak. } 
\label{tbl:Lens}
\begin{center}       
\begin{tabular}{|l|l|} 
\hline
\rule[-1ex]{0pt}{3.5ex}  Field of view of each lens & $24^\circ \times 24^\circ$ \\
\hline
\rule[-1ex]{0pt}{3.5ex}  Combined field of view & $24^\circ \times 96^\circ = 2300$~sq.~deg. \\
\hline
\rule[-1ex]{0pt}{3.5ex}  Entrance pupil diameter & 10.5~cm \\
\hline
\rule[-1ex]{0pt}{3.5ex}  Focal ratio ($f/\#$) & $f/1.4$ \\
\hline
\rule[-1ex]{0pt}{3.5ex}  Wavelength range & 600--1000~nm \\
\hline
\rule[-1ex]{0pt}{3.5ex}  Ensquared energy &
50\% within $15\times 15~\mu$m (one pixel, or $0.35\times 0.35$~arcmin) \\
\rule[-1ex]{0pt}{3.5ex}    & 90\% within $60\times 60~\mu$m ($4\times 4$ pixels, or $1.4\times 1.4$~arcmin) \\
\hline
\end{tabular}
\end{center}
\end{table} 

\subsection{Detector assembly}

The focal plane consists of four back-illuminated MIT/Lincoln~Lab
CCID-80 devices. Each CCID-80 consists of a $2048\times 2048$ imaging
array and a $2048\times 2048$ frame-store region, with 15~$\mu$m
square pixels. The frame-store region allows rapid shutterless readout
($\approx$4 ms). On each CCD there are four regions of 512 columns,
each of which has a dedicated output register. The full well capacity
is approximately 200,000 electrons. The four CCDs are abutted with a
2~mm gap, creating an effective $4096\times 4096$ imaging array
contained within a $62\times 62$~mm square (see Fig.~3).

The CCDs have an operational temperature of $-75^\circ$C to reduce the
dark current to a negligible level. They are read out at 625~kHz, and
the read noise is $<$10~$e^{-}$. The detector electronics are packaged
onto two compact double-sided printed circuit boards, each 12~cm in
diameter, which are located beneath the CCD focal plane and transmit
digitized video data over a serial LVDS link to a Data Handling Unit.

\subsection{Data Handling Unit}

The \tess Data Handling Unit (DHU) is a Space Micro Image Processing
Computer (IPC-7000) which consists of six boards: an Image Processing
Computer (IPC), which contains two Virtex-7 field-programmable gate
arrays (FPGAs) that serve as interfaces to the four cameras and
perform high-speed data processing; a Proton 400K single board
computer, which is responsible for commanding, communicating with the
spacecraft master avionics unit, and interfacing with the Ka-band
transmitter; two 192~GB solid-state buffer (SSB) cards for mass data
storage; an analog I/O power switch board to control instrument power;
and a power supply board for the DHU.

The CCDs produce a continuous stream of images with an exposure time
of 2~sec. These are received by the FPGAs on the IPC, and summed into
consecutive groups of 60, giving an effective exposure time of
2~min. During science operations, the DHU performs real-time
processing of data from the four cameras, converting CCD images into
the data products required for ground post-processing. A primary data
product is a collection of subarrays (nominally $10\times 10$ pixels)
centered on preselected target stars. The Proton400k extracts these
subarrays from each 2~min summed image, compresses them and stores
them in the SSB prior to encapsulation as CCSDS packets for the
Ka-band transmitter. Full frame images are also stacked every 30~min
and stored in the SSB. Data from the SSB are downlinked every
13.7~days at perigee.

\section{SPACECRAFT}
\label{sec:spacecraft}

\tess uses the Orbital LEOStar-2/750 bus equipped with a Ka-band
transmitter and a monopropellant (hydrazine) propulsion system. The
bus has a three-axis controlled, zero-momentum attitude control
system, and two deployed solar array wings. The total observatory
power draw is estimated to be 290~W, and the solar arrays are capable
of producing 415~W.

To achieve fine pointing, the spacecraft uses four reaction wheels and
high-precision quaternions produced by the science cameras. The
transmitter has a body-fixed high-gain antenna with a diameter of
0.7~m, a power of 2~W and a data rate of 100 Mb~s$^{-1}$. This is
sufficient to downlink the science data during 4~hr intervals at each
perigee.

The cameras are bolted onto a common plate that is attached to the
spacecraft, such that their fields of view are lined up to form a
rectangle measuring $24^\circ \times 96^\circ$ on the sky (see Figs.~4
and 7). Four elliptical holes in the plate allow shimless alignment of
the four cameras at the desired angles.

\begin{figure}[ht!]
\begin{center}
\begin{tabular}{c}
\includegraphics[height=6.5cm]{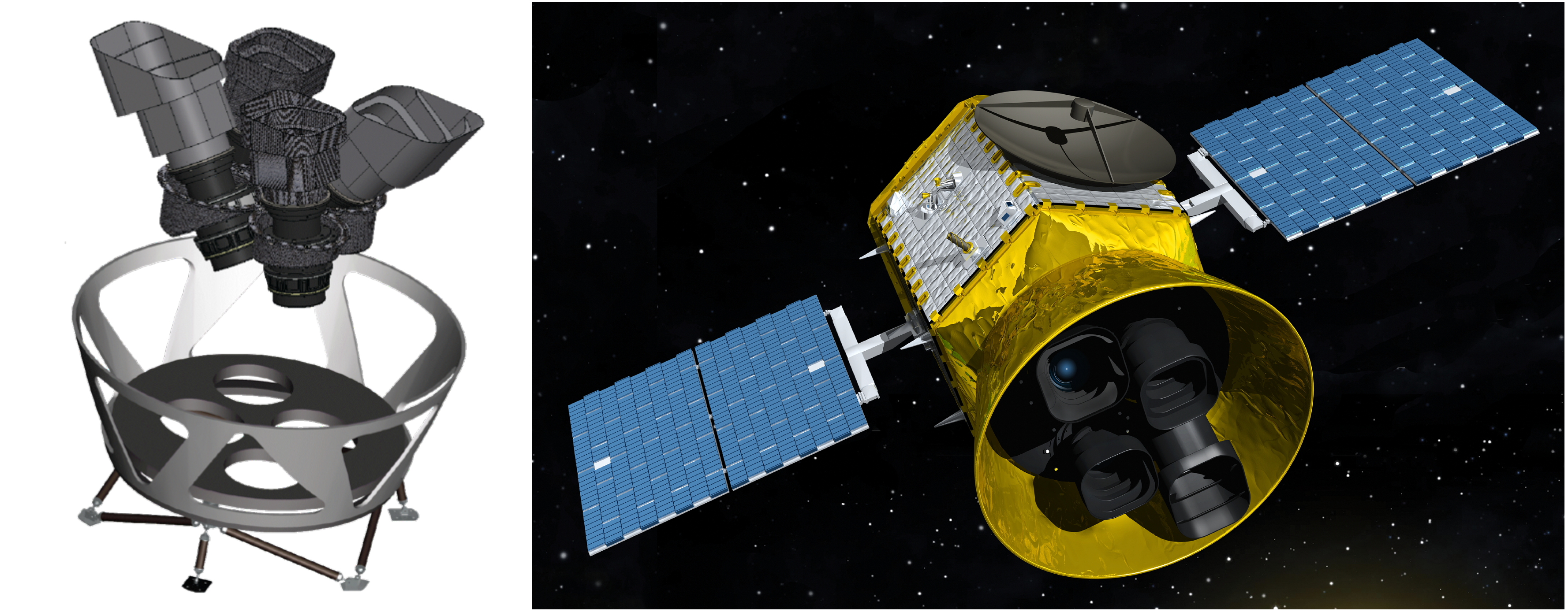}
\end{tabular}
\end{center}
\caption
{ \label{fig:GlamourShot} 
{\it Left.}---Diagram illustrating the orientations of the four
\tess cameras, lens hoods, and mounting platform. {\it
  Right.}---Artist's
conception of the \tess spacecraft and payload.}
\end{figure} 

\section{ORBIT}
\label{sec:orbit}

The \tess orbit is elliptical, with nominal perigee and apogee of
$17~R_\oplus$ and $59~R_\oplus$, and a 13.7~day period in 2:1
resonance with the Moon's orbit. The \tess orbit is inclined from the
ecliptic plane, thereby eliminating lengthy eclipses by the Earth and
Moon. The spacecraft orbit is operationally stable as a result of the
Moon leading or lagging the spacecraft apogee by $\approx$90$^\circ$,
averaging out the lunar perturbations.\cite{gangestad14} The orbit
remains above the Earth's radiation belts, leading to a relatively
low-radiation environment with a mission total ionizing dose of
$<$1~krad. The nearly constant thermal environment ensures that the
CCDs will operate near $-75^\circ$C, with temperature variations
$<$$0.1^\circ$C~hr$^{-1}$ for 90\% of the orbit, and
$<$$2^\circ$C~hr$^{-1}$ throughout the entire orbit.

This orbit can be reached efficiently using a small supplemental
propulsion system ($\Delta V \approx 3$~km~s$^{-1}$) augmented by a
lunar gravity assist. The specific path to the orbit will depend on
the launch date and launch vehicle. In a nominal scenario (illustrated
in Fig.~5), \tess is launched from Cape Canaveral into a parking orbit
with an equatorial inclination of 28.5$^\circ$. The apogee is raised
to 400,000~km after two additional burns by the spacecraft hydrazine
system, one at perigee of the first phasing orbit, and another at
perigee of the second phasing orbit. An adjustment is made at third
perigee, before a lunar flyby raises the ecliptic inclination to about
$40^\circ$. A final period-adjust maneuver establishes the desired
apogee and the 13.7 day period. The final orbit is achieved about
60~days after launch, and science operations begin soon afterward.

The orbital period and semimajor axis are relatively constant, with
long-term exchanges of eccentricity and inclination over a period of
order 8-12 years (driven by a Kozai-like mechanism\cite{gangestad14}).
There are also short-term oscillations with a period of six months
caused by solar perturbations (see Fig.~6). The orbit is stable on the
time scale of decades, or more, and requires no propulsion for
station-keeping. Table~\ref{tbl:OrbitAdvantages} lists a number of
advantages of this type of orbit for {\it TESS}.

\begin{figure}[ht!]
\begin{center}
\begin{tabular}{c}
\includegraphics[height=8cm]{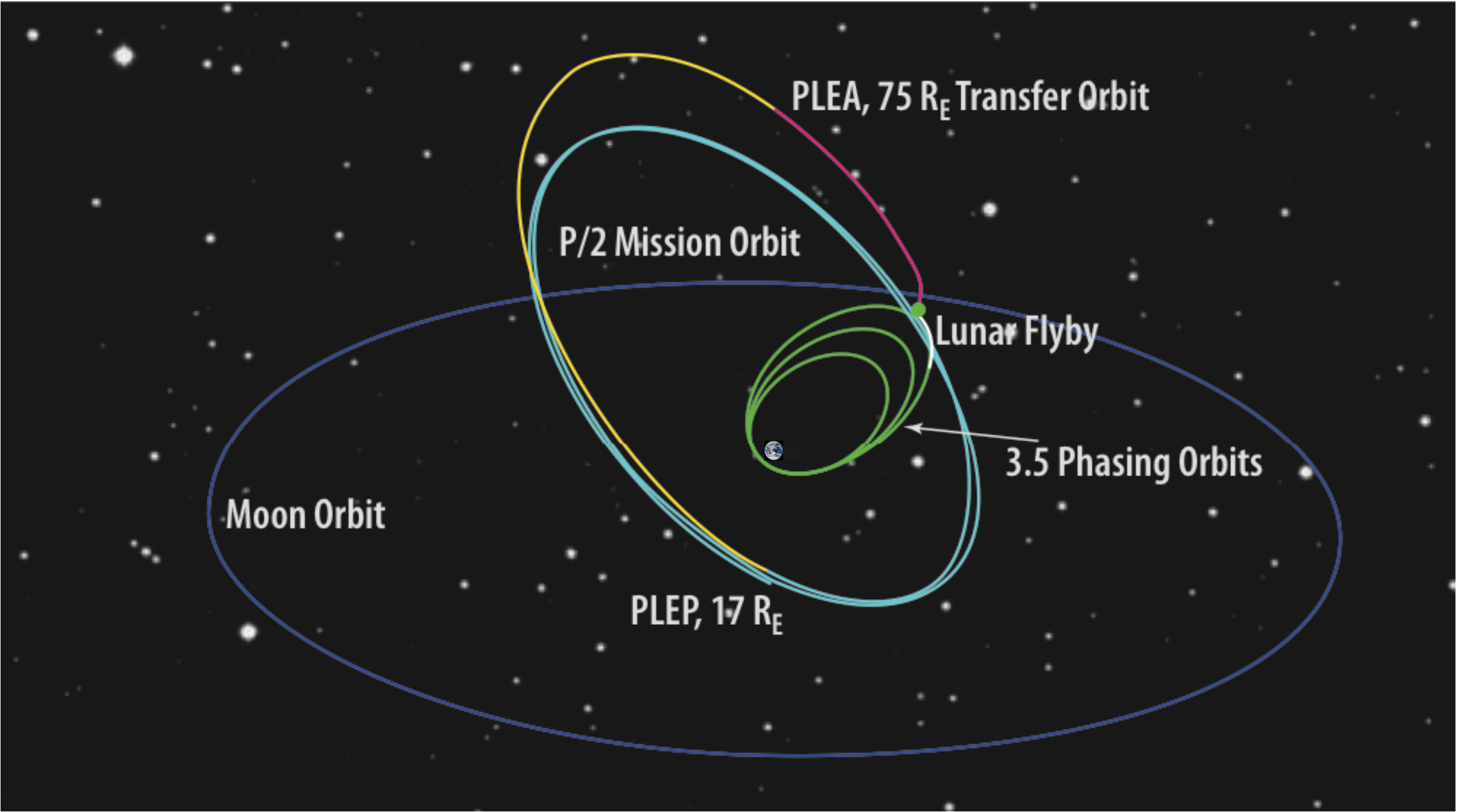}
\end{tabular}
\end{center}
\caption
{ \label{fig:Orbit} 
Maneuvers and scenario for achieving the \tess mission orbit.
PLEP is post--lunar-encounter perigee, and PLEA is
post--lunar-encounter apogee.
}
\end{figure} 

\begin{table}[hb!]
\caption{Characteristics of the \tess spacecraft orbit and comparisons
  to a low-Earth orbit. } 
\label{tbl:OrbitAdvantages}
\begin{center}       
\begin{tabular}{|l|}
\hline
\rule[-1ex]{0pt}{3.5ex}  Extended and unbroken observations: $>$300~hr~orbit$^{-1}$ \\
\hline
\rule[-1ex]{0pt}{3.5ex}  Thermal stability: $<$0.1$^\circ$C~hr$^{-1}$ (passive
control) \\
\hline
\rule[-1ex]{0pt}{3.5ex}  Earth/Moon stray light: $\sim$$10^6$ times lower
than in low-Earth orbit \\
\hline
\rule[-1ex]{0pt}{3.5ex}  Low radiation levels: no South Atlantic
anomaly or outer belt electrons \\
\hline
\rule[-1ex]{0pt}{3.5ex}  Frequent launch windows: 20 days per lunation \\
\hline
\rule[-1ex]{0pt}{3.5ex}  High data rates at perigee: $\sim$100~Mb~s$^{-1}$ \\
\hline
\rule[-1ex]{0pt}{3.5ex}  Excellent pointing stability: no drag or
gravity gradient torques\\
\hline
\rule[-1ex]{0pt}{3.5ex}  Simple operations: single 4-hr downlink \& 
repoint every 2 weeks \\
\hline
\rule[-1ex]{0pt}{3.5ex}  Long lifetime: several decades above GEO ($>$6.6~$R_\oplus$) \\
\hline
\end{tabular}
\end{center}
\end{table} 

\begin{figure}
\begin{center}
\begin{tabular}{c}
\includegraphics[height=8cm]{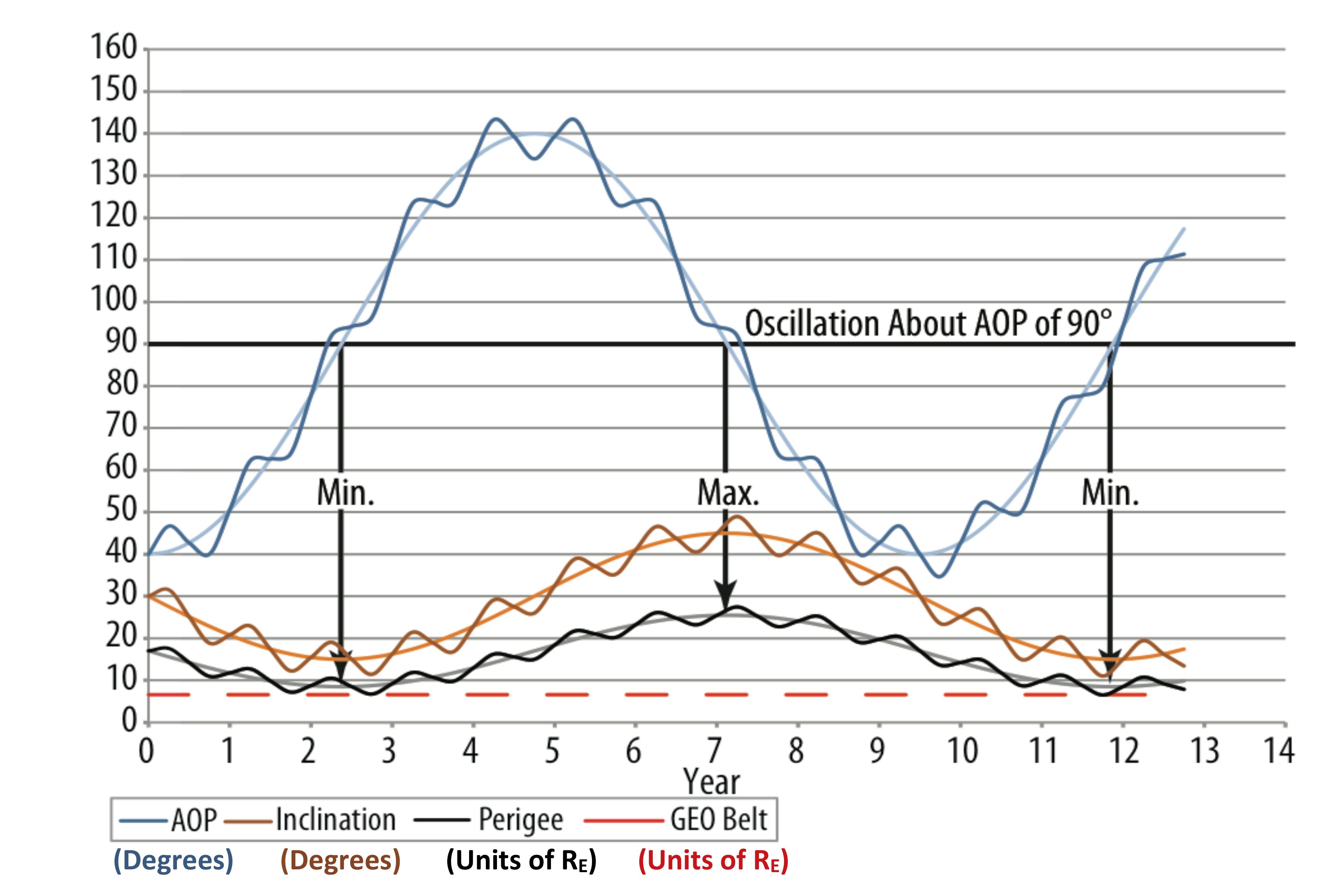}
\end{tabular}
\end{center}
\caption
{ \label{fig:OrbitPerturbations} 
Calculated time variations in the elements of the nominal
\tess mission orbit. The units of each curve are specified
in the legend. AOP is the argument of perigee. GEO is geosynchronous
Earth orbit.}
\end{figure} 

\section{SURVEY OPERATIONS}
\label{sec:survey}

\subsection{Star selection}

A catalog of preselected stars will be monitored at a cadence of 2~min
or better. Ideally the catalog will include $\gsim$200,000
main-sequence FGKM stars that are sufficiently bright and small to
enable the detection of transiting planets smaller than Neptune. This
corresponds to approximately $I_C <12$ for FGK stars and $I_C<13$ for
M dwarfs.\cite{sullivan14} These stars are bright enough that it
should be possible to assemble a list using currently available
all-sky surveys. A draft catalog has been constructed using the
2MASS\cite{skrutskie06}, Tycho-2\cite{hog00}, and
UCAC4\cite{zacharias13} catalogs, supplemented with smaller catalogs
of nearby M dwarfs. For the M stars, infrared magnitudes allow for
discrimination between dwarfs and giants.\cite{bessell88} G and K
dwarfs are distinguished from giants through their relatively rapid
reduced proper motion.\cite{salim02} It may also be possible to use
trigonometric parallaxes measured by the ongoing European {\it Gaia}
mission, if they become available early enough for \tess mission
planning. The \tess target catalog will be publicly released 
prior to launch.

\subsection{Scanning strategy}

\begin{figure}
\begin{center}
\begin{tabular}{c}
\includegraphics[height=7cm]{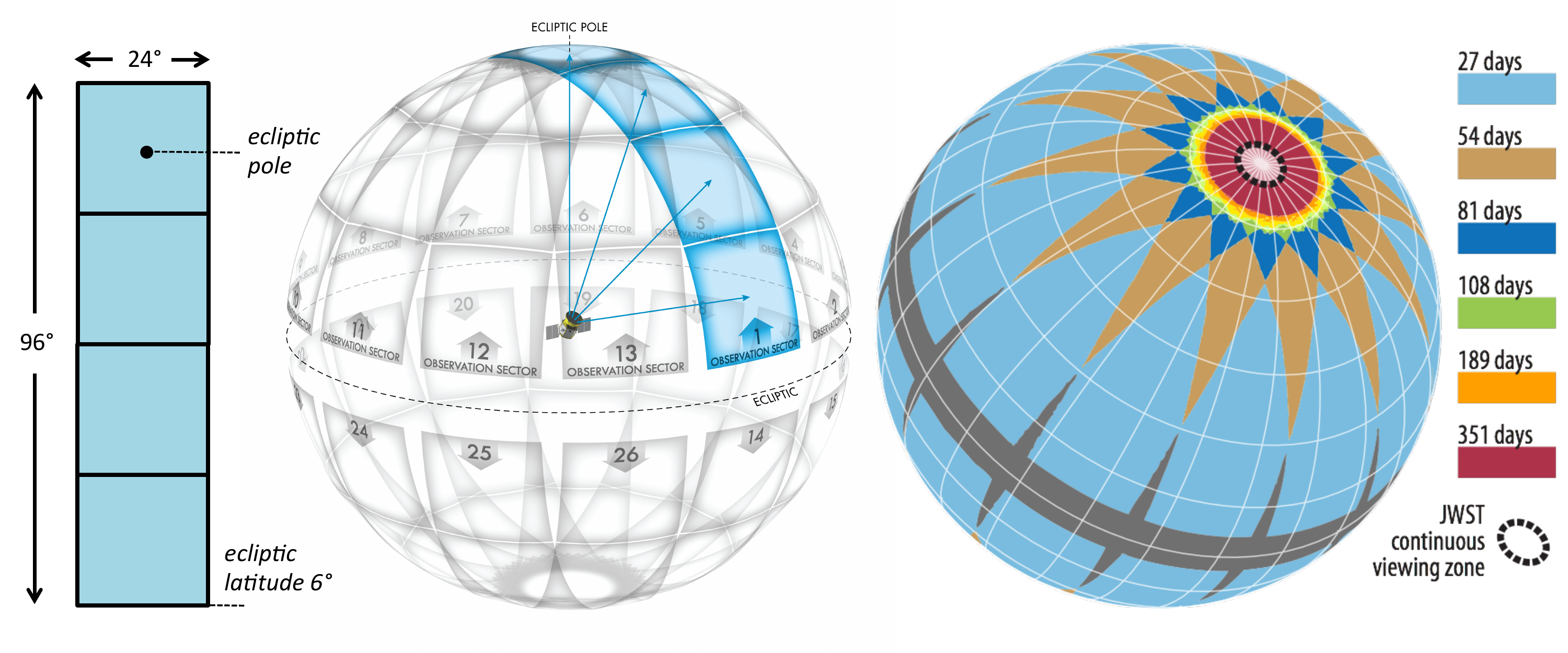}
\end{tabular}
\end{center}
\caption
{ \label{fig:SkyCoverage} 
{\it Left.}---The instantaneous combined field of view of the four
\tess cameras. 
{\it Middle.}---Division of the celestial sphere into 26 observation sectors (13
per hemisphere).
{\it Right.}---Duration of observations on the celestial sphere,
taking into account the overlap between sectors.
The dashed black circle enclosing the ecliptic pole shows the
region which {\it JWST} will be able to observe at any time.}
\end{figure} 

The four cameras act as a $1\times 4$~array, providing a combined FOV
of $24^\circ \times 96^\circ$ or 2300 square degrees (see Fig.~7). The
north and south ecliptic hemispheres are each divided into 13
partially overlapping sectors of $24^\circ \times 96^\circ$, extending
from an ecliptic latitude of 6$^\circ$ to the ecliptic pole. Each
sector is observed continuously for two spacecraft orbits (27.4~days),
with the boresight of the four-camera array pointed nearly
anti-solar. After two orbits, the FOV is shifted eastward in ecliptic
longitude by about $27^\circ$, to observe the next sector. Observing
an entire hemisphere takes one year, and the all-sky survey takes two
years.\footnote{A video illustrating the \tess survey strategy, along
  with the pathway to the spacecraft orbit, can be seen at {\tt
    http://www.youtube.com/watch?v=mpViVEO-ymc}.}

The overlap of the sectors is illustrated in Fig.~7. Approximately
30,000 square degrees are observed for at least 27 days.  Close to the
ecliptic poles, approximately 2800 square degrees are observed for
more than 80 days. Surrounding the ecliptic poles, approximately 900
square degrees are observed for more than 300 days.

\subsection{Data Downlink and Housekeeping Operations}

At perigee, science operations are interrupted for no more than 16
hours to point {\it TESS}\,'s antenna toward Earth, downlink data, and
resume observing.  This includes a nominal 4~hr period for Ka-band
science data downlink using NASA's Deep Space Network.  In addition,
momentum unloading is occasionally needed due to the $\approx$1.5 N~m
of angular momentum build-up induced by solar radiation pressure. For
this purpose \tess uses its hydrazine thrusters.

\subsection{Ground-based data analysis and follow-up} 

The \tess data will be processed with a data reduction pipeline based
on software that was developed for the {\it Kepler}
mission.\cite{jenkins10} This includes pixel-level calibration,
background subtraction, aperture photometry, identification and
removal of systematic errors, and the search for transits with a
wavelet-domain matched filter.

Once the data are processed and transits are identified, selected
stars will be characterized with ground-based imaging and
spectroscopy.  These observations are used to establish reliable
stellar parameters, confirm the existence of planets, and establish
the sizes and masses of the planets. Observations will be performed
with committed time on the Las Cumbres Observatory Global Telescope
Network and the MEarth observatory. In addition the \tess science team
members have access to numerous other facilities (e.g., Keck,
Magellan, Subaru, HARPS, HARPS-North, Automated Planet Finder) through
the usual telescope time allocation processes at their home
institutions. The \tess team includes a large group of collaborators
for follow-up observations and welcomes additional participation.

\section{ANTICIPATED RESULTS} 
\label{sec:results}

\subsection{Photometric performance}

\begin{figure}
\begin{center}
\begin{tabular}{c}
\includegraphics[height=9cm]{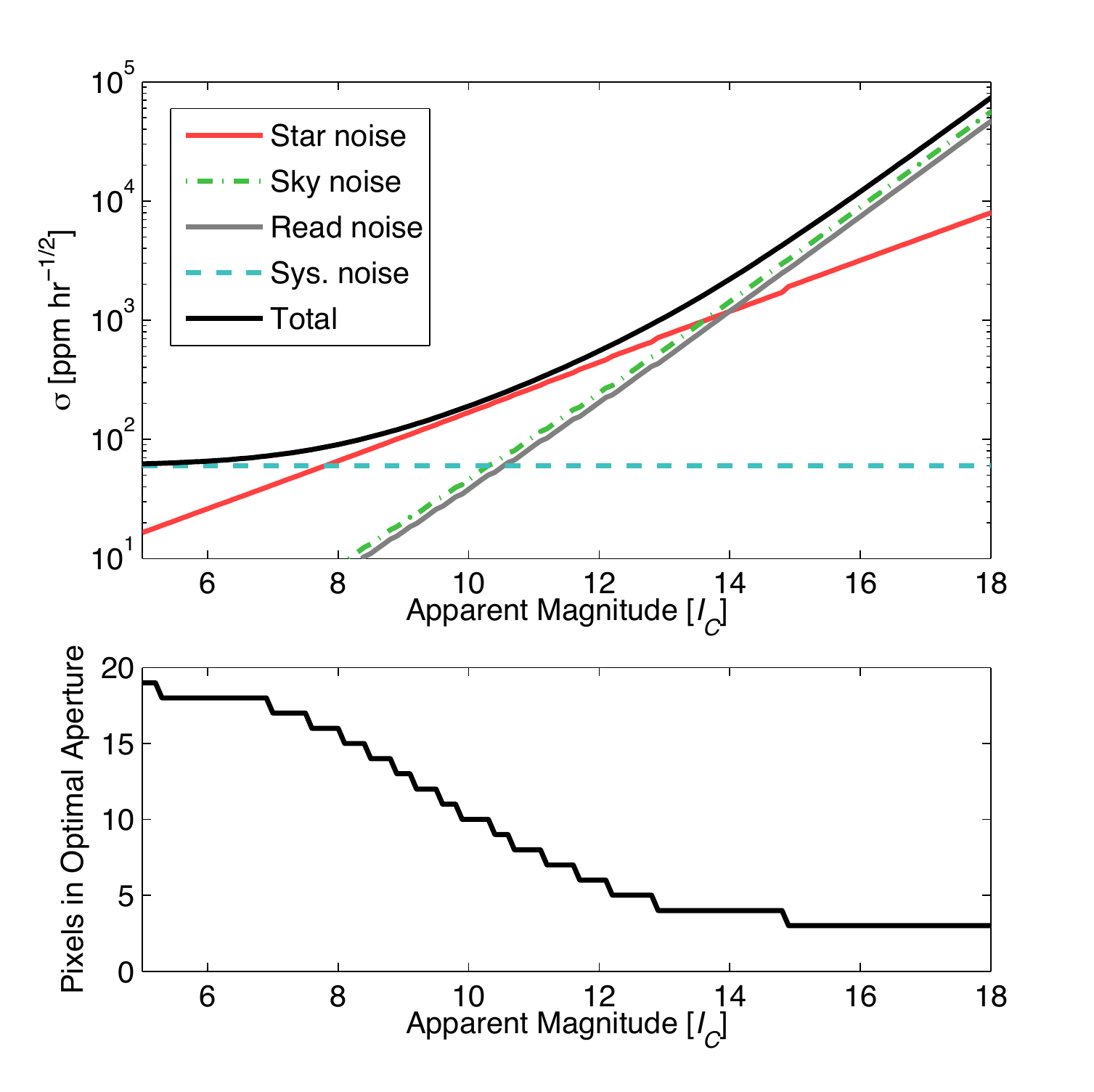}
\end{tabular}
\end{center}
\caption
{ \label{fig:Noise} 
{\it Top.}---Expected 1$\sigma$ photometric precision as a function
of stellar apparent magnitude in the $I_C$ band.
Contributions are from photon-counting noise from the target
star and background (zodiacal light and unresolved stars),
detector read noise (10~$e^{-}$), and
an assumed 60~ppm of incorrigible noise on hourly timescales.
{\it Bottom.}---The number of pixels in the photometric aperture that
optimizes the signal-to-noise ratio.}
\end{figure} 

Figure~8 shows the anticipated photometric performance of the \tess
cameras. The noise sources in this model are photon-counting noise
from the star and the background (zodiacal light and faint unresolved
stars), dark current (negligible), readout noise, and a term
representing additional systematic errors that cannot be corrected by
co-trending. The most important systematic error is expected to be due
to random pointing variations (``spacecraft jitter''). Because of the
non-uniform quantum efficiency of the CCD pixels, motion of the star
image on the CCD will introduce changes in the measured brightness, as
the weighting of the image PSF changes, and as parts of the image PSF
enter and exit the summed array of pixels.

The central pixel of a stellar image will saturate at approximately
$I_C=7.5$. However, this does not represent the bright limit for
precise photometry because the excess charge is spread across other
CCD pixels and is conserved, until the excess charge reaches the
boundary of the CCD. As long as the photometric aperture is large
enough to encompass all of the charge, high photometric precision can
still be obtained. The {\it Kepler} mission demonstrated that
photon-noise--limited photometry can be obtained for stars 4~mag
brighter than the single-pixel saturation limit\cite{gilliland10}.
Since similar performance is expected for {\it TESS}, the bright limit
is expected to be $I_C\approx 4$ or perhaps even brighter.

\subsection{Transit detections}

Monte Carlo simulations are used to verify that the science objectives
can be met and to anticipate the properties of the detected planetary
systems.\cite{sullivan14} These simulations are based on a model of
the local neighborhood of main-sequence FGKM stars.\cite{girardi05}
Simulated stars are populated randomly with planets, and ``observed''
by {\it TESS}. Those for which transits are observed with a
sufficiently high signal-to-noise ratio are counted as detections. In
addition, the simulated star catalog is populated with eclipsing
binaries that may be blended with brighter stars to produce
transit-like signals; the detections of these ``false positives'' are
also recorded.

Among the features of the current simulations are: (1) a realistic
distribution of stars and eclipsing binaries based on local census
data; (2) probability distributions for planetary occurrence and
orbital properties taken from the {\it Kepler}
results\cite{fressin13}; (3) variation in stellar surface density and
zodiacal light with position on the celestial sphere; (4) variation in
the duration of \tess observations depending on ecliptic coordinates.

Figure~9 illustrates some of the results. \tess is expected to find
thousands of planets smaller than Neptune, including hundreds of
super-Earths (1.25--2~$R_\oplus$) and tens of planets comparable in
size to Earth. These will be accompanied by a comparable number of
false positives (as has been the case for the {\it Kepler} mission), a
majority of which are background eclipsing binaries.  Some false
positives will be identifiable using \tess data alone, based on the
detection of secondary eclipses, ellipsoidal flux modulation, or
transit-specific image motion. In other cases ground-based
observations will be required to check for composite spectra, large
radial-velocity variations, color-dependent transit depths, and
resolved companions that are indicative of false positives.

\begin{figure}
\begin{center}
\begin{tabular}{c}
\includegraphics[height=11cm]{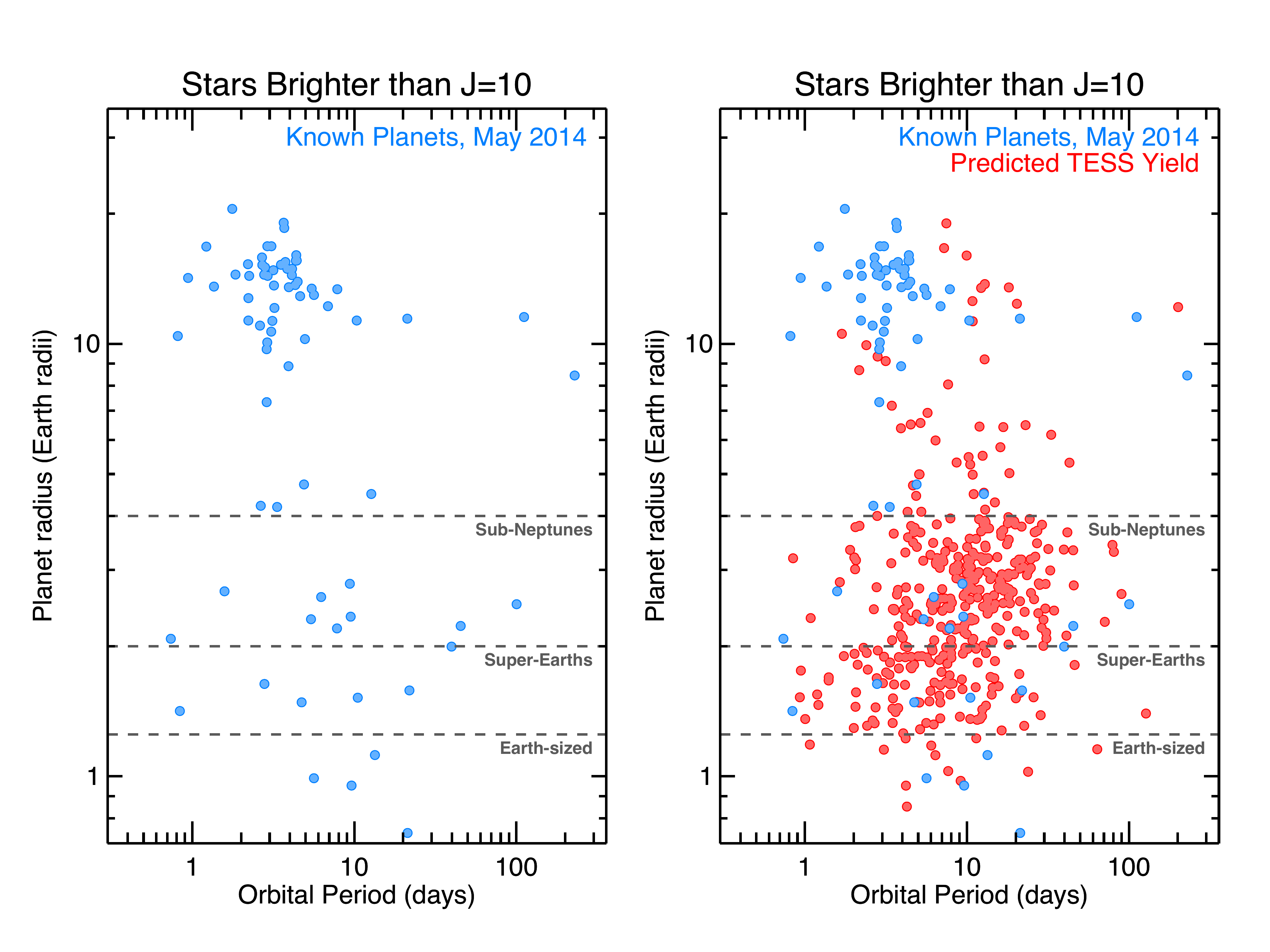}
\end{tabular}
\end{center}
\caption
{ \label{fig:BeforeAfter} 
Sizes and orbital periods of planets with host stars brighter than
$J=10$. The $J$ band was chosen for convenience, since the 2MASS survey
provides $J$ magnitudes for all of the known
planet-hosting stars.
{\it Left.}---Currently known planets, including those from
the {\it Kepler} and {\it CoRoT} missions as well as ground-based surveys.
{\it Right}.---Including the simulated population of \tess
detections.}
\end{figure} 

\subsection{Asteroseismology}
\label{sec:asteroseismology}

Observing photometric variations due to stellar oscillations
(``asteroseismology'') provides sensitive diagnostics of the stellar
mass, radius, and internal dynamics. Based on the {\it Kepler}
experience with mode amplitudes as a function of stellar
parameters\cite{chaplin10}, \tess can be expected to detect $p$-mode
oscillations on about 6,000 stars brighter than $V = 7.5$, including
(a) the majority of all stars brighter than $V = 4.5$, (b) about 75
stars smaller than the Sun, (c) about 2,000 upper-main-sequence and
subgiant stars, and (d) virtually all the giant stars. Stars that are
not suitable for planet searching but are appropriate for
asteroseismology (such as giants) can be added to the \tess input
catalog at minimal cost.

The key features enabling {\it TESS}'s advances in this area are
all-sky coverage and relatively fine time sampling. Compared to
similar stars studied by {\it CoRoT} and {\it Kepler}, the \tess stars
will be more numerous, brighter, and nearer to the Earth. \tess stars
will have accurately known parallaxes, will be much more amenable to
interferometric and radial-velocity studies, and are themselves among
the likely targets for future, imaging-based missions to study
exoplanets.

\subsection{Full Frame Images}

In addition to monitoring 200,000 pre-selected stars with a 2~min
cadence, {\it TESS} will return a nearly continuous series of full
frame images (FFI) with an effective exposure time of 30~min. As {\it
  Kepler} has shown, a 30~min cadence still allows for the efficient
discovery of transiting planets. The FFIs will expand the transit
search to any star in the fields of view that is sufficiently bright,
regardless of whether it was selected ahead of time. This will reduce
the impact of any imperfections in the target star catalog, and allow
the search sample to extend beyond the FGKM dwarfs that are the focus
of the mission. In addition, the transit candidates that are
identified in the FFIs during the baseline {\it TESS} mission could be
selected for shorter-cadence observations during an extended mission.

The FFIs will also enable a wide variety of other scientific
investigations, such as the detection and monitoring or nearby
supernovae and gamma-ray burst afterglows, bright AGN outbursts,
near-Earth asteroids, eclipsing and close binaries, novae, flaring
stars, and other variable stars. Each 30~min FFI will provide precise
($\approx$5~mmag) photometry for approximately $10^6$ bright galaxies
and stars ($I_C <14$-15) within a 2300~deg$^2$ field. Over two
years, the \tess all-sky survey will provide such data for
approximately 20 million bright objects, many of which will exhibit
short-term variability. Thus, the \tess FFIs will provide
broad-ranging fundamental data for time-domain astronomy.  In
particular, \tess will complement the Large Synoptic Survey Telescope,
which is limited to observing objects fainter than 16th magnitude
\cite{lsst}.

\section{PARTNERS AND THEIR ROLES} 
\label{sec:partners}

The lead institution for \tess is the Massachusetts Institute of
Technology (MIT), which hosts the Principal Investigator, Dr.\ George
Ricker. The MIT Lincoln Laboratory is responsible for the cameras,
including the lens assemblies, detector assemblies, lens hoods, and
camera mount. NASA's Goddard Space Flight Center provides project
management, systems engineering, and safety and mission
assurance. Orbital Sciences Corporation (OSC) builds and operates the
spacecraft. The mission is operated from the OSC Mission Operations
Center.

The \tess Science Center, which analyzes the science data and
organizes the co-investigators, collaborators, and working groups
(with members from many institutions) is a partnership among MIT's
Physics Department and Kavli Institute for Astrophysics and Space
Research, the Smithsonian Astrophysical Observatory, and the NASA Ames
Research Center. The raw and processed data are archived at the
Mikulski Archive for Space Telescopes (MAST), at the Space Telescope
Science Institute.

\section{DISCUSSION} 

{\it Kepler} has revealed the vast numbers of small planets that exist
in the Galaxy. \tess will find them around stars in the local
neighborhood, which will be easier to study with current and planned
instruments, including the {\it JWST}. The \tess launch is currently
scheduled for late 2017, and the all-sky survey for 2018-2019. An
extended mission of several years is a realistic possibility, and is
compatible with the supply of consumable materials and the stability
of the spacecraft orbit. An extended mission would increase the number
of planet detections, especially those with longer orbital periods,
including planets in the habitable zones of their host stars.

A broad community effort will be needed to harvest all of the
scientific potential of the survey. \tess will serve as the ``People's
Telescope,'' with a public release of the target catalog prior to
launch, the first data release 6 months after launch, and subsequent
data releases every 4 months, to stimulate community-wide effort and
optimize target selection for subsequent studies. Light curves,
full-frame images, and catalogs of ``objects of interest'' will be
available on the MAST. The \tess team will also establish an
electronic clearinghouse to exchange information and coordinate the
follow-up effort.

One of the most scientifically productive---and challenging---types of
follow-up observations is precise radial-velocity (PRV) monitoring.
PRV monitoring is used to confirm the planetary nature of \tess
detections, provide constraints on planetary densities and possible
interior compositions, and inform the interpretation of planetary
atmosphere studies with {\it JWST} and other instruments. PRV
monitoring is likely to be the rate-limiting step in realizing the
full benefits of \tess for exoplanetary science.  Even though \tess
stars are relatively bright and favorable for study, the expected
number of new and interesting PRV targets will greatly exceed the time
budgets of currently available observatories. Many groups are aiming
to rectify this situation by building new PRV instruments. Among the
new instruments being planned with {\it TESS} targets in mind are
ESPRESSO\cite{pepe2010}, NRES\cite{eastman2014}, IRD\cite{tamura2012},
CARMENES\cite{quirrenbach2010}, HPF\cite{mahadevan2012},
HRS\cite{bramall2012}, SPIRou\cite{delfosse2013},
MAROON-X\cite{bean2014}, CODEX\cite{pasquini2008}, and
G-CLEF\cite{szentgyorgyi2012}.

The \tess legacy will be a catalog of the nearest and brightest stars
hosting transiting planets, which will likely be the most favorable
targets for detailed investigations in the decades or even centuries
that follow.

\acknowledgments Many people and institutions have generously
supported \tess over the years, including:
Aerospace Corporation,
Google,
the Kavli Foundation,
the MIT Department of Physics,
the MIT School of Science,
Mr.\ Gregory E.\ Moore and
Dr.\ Wynne Szeto,
Mr.\ Richard M.\ Tavan,
and Mr.\ Juan Carlos Torres.
Extensive support has also been provided
by NASA Headquarters, NASA's Goddard Space Flight Center, and NASA's
Ames Research Center (ARC) under the following grants and contracts: NNG09FD65C,
NNX08BA61A, NNG12FG09C, and NNG14FC03C.
The authors also
wish to thank the following individuals for their important scientific,
technical, and other contributions to the mission:
Charles Alcock,
Fash Asad, 
Mark Bautz,
Chet Beals,
Dave Bearden,
Marc Bernstein,
Greg Berthiaume,
Ed Bertschinger,
Adam Burgasser,
Barry Burke,
Claude Canizares,
Ben Cichy,
Kris Clark,
Dave Czajkowski,
Debra Emmons,
Jim Francis,
Joe Gangestad,
Bob Goeke,
Jose Guzman,
Kari Haworth,
Greg Henning,
Jackie Hewitt,
Shane Hynes,
Marc Kastner,
Brian Lewis,
Robert Lockwood,
Gerry Luppino,
Francois Martel,
Bill Mayer,
Chad Mendelsohn,
Ed Morgan,
Bill Oegerle,
Randy Persinger,
Ron Remillard,
Matt Ritsko,
Tim Sauerwein,
Robbie Schingler,
Joe Scillieri,
Rob Simcoe,
Tony Smith,
Dave Strobel,
Vyshi Suntharalingam,
Jeff Volosin,
Kim Wagenbach,
Nick White,
Pete Worden,
and Maria Zuber.


\bibliography{report}   
\bibliographystyle{spiejour}   


\vspace{2ex}\noindent{\bf Dr.\ George Ricker} is the PI for the
Transiting Exoplanet Survey Satellite mission. He is a Senior Research
Scientist and Director of the CCD Laboratory at the MIT Kavli
Institute. Dr.\ Ricker was the PI for the HETE mission (launched in
2000), PI for the CCD X-ray camera on the ASCA mission (1993),
Deputy-PI for the Chandra ACIS instrument (1999), and the US PI for
the X-ray CCD Camera on the Astro-E1 mission.

\vspace{1ex}
\noindent Biographies and photographs of the other authors are not available.

\listoffigures

\listoftables

\end{spacing}
\end{document}